\documentclass[11pt, oneside]{article}   	
\usepackage[margin=1in]{geometry}               
\usepackage{graphicx}				
										
\usepackage{amssymb}
\usepackage{physics}
\usepackage{amsmath}
\usepackage[dvips]{epsfig}

\usepackage{amsfonts}
\usepackage{subfig}

\usepackage{mathrsfs}

\usepackage{bm}

\usepackage[title]{appendix}

\usepackage{cite}

\usepackage{authblk}

\def\cchi{\raise2pt\hbox{$\chi$}} 

\newcommand\bsdot{\ensuremath{\boldsymbol{.}}}

\title{\bf{The Effect of the Pauli Spin Matrices on the Quantum Lattice Algorithm for Maxwell Equations in Inhomogeneous Media}}

\author{George Vahala}
\affil{Department of Physics, William \& Mary, Williamsburg, VA23185}
\author{Linda Vahala}
\affil{Department of Electrical \& Computer Engineering, Old Dominion University, Norfolk, VA 12319}
\author{Min Soe}
\affil{Department of Mathematics  and Physical Sciences, Rogers State University, Claremore, OK 74017}
\author{Abhay K. Ram}
\affil{Plasma Science and Fusion Center, MIT, Cambridge, MA 02139} 

\begin{document}
\maketitle
$\bf{Abstract}$:  A quantum lattice algorithm (QLA) is developed for the solution of Maxwell equations in scalar dielectric media using the Riemann-Silberstein representation.
For x-dependent and y-dependent inhomogeneities, the corresponding QLA requries 8 qubits/spatial lattice site.  This is because the corresponding Pauli spin matrices have off-diagonal components which  permit the collisional entanglement of two qubits.  However, z-dependent inhomogeneities require a QLA with 16 qubits/lattice site since the Pauli spin matrix $\sigma_z$ is diagonal.  QLA simulations are performed for the time evolution of an initial electromagnetic pulse propagating normally to a boundary layer region joining two media of different refractive index. There is excellent agreement between all three representations, as well as very good agreement with nearly all the standard plane wave boundary condition results for reflection and transmission off a dielectric discontinuity.  In the QLA simulation, no boundary conditions are imposed at the continuous, but sharply increasing, boundary layer.


\section{Introduction}
Quantum lattice algorithms (QLA) [1-20] have been shown to be excellent perturbative representations of physical problems that are ideally parallelized
on classical supercomputers, and able to be directly encoded on a quantum computer.  QLAs consist of an interleaved sequence of unitary collision and streaming operators on a set of qubits.  The collision and streaming operators do not commute.  Typically this part of QLA will recover the differential structures of the physical problem under consideration.  For physics problems like the Nonlinear Schrodinger equation (NLS), the nonlinear terms are handled perturbatively by an exponential potential operator.  The ensuing QLA [9-11] accurately recovered all the physics of soliton collisions, including the signature phase change induced by the actual collisions themselves.  The extension of the NLS equation to three dimensions (3D) now permits the examination of the ground state of a Bose Einstein condensate (BEC) using QLA [4, 12-16, 19].  This has  led to quantum turbulence studies and vortex reconnection in scalar and spinor BECs.

Here we continue our studies of a QLA for Maxwell equations in inhomogeneous media.  In our earlier papers [17, 31] we presented QLA for 1D propagation of electromagnetic fields in a scalar dielectric medium. These QLAs were based on the Riemann-Silberstein vectors, which in essence give the two polarizations of an electromagnetic pulse.  For homogeneous dielectrics, there is remarkable similarity between the Dirac equation and the 4-vector Riemann-Silberstein representation of Maxwell equations.  Thus the Pauli spin-$1/2$ matrices play a significant role in the QLA.  Khan [21] showed that for inhomogeneous media,  the terms proportional to the gradient of the refractive index $n'(x)$ will lead to non-unitary operators in the time evolution of Maxwell equations.  When one determines a QLA for Maxwell equations in an inhomogeneous medium, some of the evolution operators will necessarily be Hermitian, rather than unitary.  In particular, for 1D propagation in the y-direction [17] two of the evolution operators are Hermitian, while for 1D x-propagation only one of the operators is Hermitian.   Interestingly, Childs and Wiebe [22] have shown that algorithms utilizing sum of unitary operators (rather than the standard product of unitary operators) can still be encoded onto a quantum computer.  

In [17] we  discussed how to determine a 1D QLA for Maxwell equations with a y-dependent dielectric and in [31] that for an x-dependent dielectric.  Here we discuss how to develop 1D QLA for Maxwell equations for  the z-dependent refractive indices.  Once having determined these three orthogonal 1D QLA representations one can immediately stitch these representations together to develop both 2D and fully 3D QLA representations of Maxwell equations.   For both the x-dependent and y-dependent dielectrics, an 8-qubit representation is sufficient.  However, for z-dependent dielectrics one will require a 16-qubit representation.  This difference between these representations  arises from the fact that the z-component Pauli spin matrix $\sigma_z$ is diagonal.  The collision operator requires the coupling of at least 2 qubiits locally at each lattice site in order to get entanglement.  This entanglement is then spread throughout the lattice by the streaming operators.

Khan's representation [21] of the Maxwell equations in an inhomogeneous medium using the Riemann-Silberstein vectors is presented in Sec. 2.  In Sec. 3 we present the QLA formulation for x-, y-, and z-dependent media, while simulation results for z- dependent media are given in Sec. 4.  Once one has these 1D modular representations, the QLA for Maxwell equations with either 2D or 3D dielectric inhomogeneities  can be readily determined.  Some QLA preliminary simulation results for 2D Maxwell equations are presented in Sec. 5.

\section{General Theory of Khan [21] for Maxwell equations in Inhomogeneous Media}

Soon after Dirac [23] was able to determine the square root of the  
Klein-Gordon wave operator and thus obtain a relativistically invariant counterpart  
to the Schr\"odinger equation, interest developed in making a formal theoretical
connection between the relativistically invariant Maxwell equations and the 
Dirac equation [24-28]. One particularly intriguing approach has been through
the use of the Riemann-Silberstein vectors [21, 24]
\begin{equation}
\label{R-S vector}
\mathbf{F^{\pm}} = \sqrt{\epsilon} \mathbf{E}  \pm i \frac{\mathbf{B}}{\sqrt{\mu}}.
\end{equation}
where $\mathbf{E}$ is the (real) electric field, $\mathbf{B}$  the magnetic flux density, and
$\epsilon$ and $\mu$ are the (scalar) permittivity and permeability of the medium, respectively.
Thus the electric displacement $\mathbf{D} = \epsilon \mathbf{E}$, and the magnetic field 
$\mathbf{H} = \mathbf{B}/ \mu$ .  The Maxwell equations (with free charge density $\rho$ and free current density 
$\mathbf{J}$)
\begin{align}
\nabla \bsdot \mathbf{D}  \ &= \ \rho \ & \ \nabla \bsdot \mathbf{B} \ &= \ 0 \label{gauss} \\
\nabla \times \mathbf{E} \ &= \ - \frac{\partial \mathbf{B}}{\partial t}  \ & 
\ \nabla \times \mathbf{H} \ &= \ \mathbf{J} + \frac{\partial \mathbf{D}}{\partial t} 
\label{fam}
\end{align}
can be written in the Riemann-Silberstein form [21],
\begin{equation}
i \frac{\partial \mathbf{F^{\pm}}}{\partial t} = \pm v \nabla \times \mathbf{F^{\pm}} 
\pm \frac{1}{2} \nabla v \times \mathbf{F^{\pm}} \pm \frac{v}{2 h} \nabla h \times
\mathbf{F^{\mp}} + \frac{i}{2} \left( \frac{\partial \ln v}{\partial t} \mathbf{F^{\pm}} +
\frac{\partial \ln h}{\partial t} \mathbf{F^{\mp}} \right) - i \sqrt{\frac{v h}{2}} 
\mathbf{J},
\end{equation}
\begin{equation}
\nabla \bsdot \mathbf{F^{\pm}} = \frac{1}{2 v} \nabla v \bsdot \mathbf{F^{\pm}} +
\frac{1}{2 h} \nabla h \bsdot \mathbf{F^{\pm}} + \sqrt{\frac{v h}{2}} \rho .
\end{equation}
where $v$ is the normalized electromagnetic phase velocity of the wave in the medium, and
$h$ is a normalized resistance
\begin{equation}
v = \frac{1}{\sqrt{\epsilon \mu}},\ \ \ \   h = \sqrt{\frac{\mu}{\epsilon}}.
\end{equation}
Note that the coupling between the two field polarizations $\mathbf{F^{\pm}}$ occurs through 
the space-time variations in $h$.  

In determining the QLA representation of Maxwell equations it is convenient to rewrite the system in matrix form [21],,
\begin{equation}
\begin{aligned}
\renewcommand\arraystretch{1.8}
\frac{\partial}{\partial t} \begin{pmatrix} \Psi^+ \\ \Psi^- \end{pmatrix}
- & \renewcommand\arraystretch{1.8}
\frac{1}{2} \frac{\partial \ln v}{\partial t} \begin{pmatrix} \Psi^+ \\ \Psi^- \end{pmatrix}
+\frac{i}{2} M_z \alpha_y \frac{\partial \ln h}{\partial t}
\begin{pmatrix} \Psi^+ \\ \Psi^- \end{pmatrix}
=  \\ 
& \renewcommand\arraystretch{2.3}
 - v
\begin{pmatrix} \mathbf{M} \bsdot \nabla + \boldsymbol{\Sigma} \bsdot 
\displaystyle{\frac{\nabla \nu}{2 \nu}}
& -i M_z \boldsymbol{\Sigma} \bsdot \displaystyle{\frac{\nabla h}{h}} \alpha_y \\
-i M_z \boldsymbol{\Sigma}^{*} \bsdot \displaystyle{\frac{\nabla h}{h}} \alpha_y &
\mathbf{M}^{*} \bsdot \nabla + \boldsymbol{\Sigma}^{*} \bsdot 
\displaystyle{\frac{\nabla \nu}{2 \nu}} \end{pmatrix}
\begin{pmatrix} \Psi^+ \\ \Psi^- \end{pmatrix} -
\begin{pmatrix} W^+ \\ W^- \end{pmatrix},
\end{aligned}
\end{equation}
where the 8-spinor Cartesian components of the Riemann-Silberstein vectors are,
\begin{equation}
\begin{aligned}
\renewcommand\arraystretch{1.6}
{\Psi^{\pm} }
=  
\begin{pmatrix}
- F^{\pm}_x \pm i F^{\pm}_y    \\
F^{\pm}_z \\
F^{\pm}_z \\
 F^{\pm}_x \pm i F^{\pm}_y  
\end{pmatrix}
.
\end{aligned}
\end{equation}
The $4 \times 4$ matrices $\mathbf{M}$ are the tensor products 
of the Pauli spin matrices,
$\boldsymbol{\sigma} = (\sigma_x , \sigma_y , \sigma_z) $
\begin{equation}
\label{Pauli_spin}
\sigma_x = 
\begin{pmatrix} 
0 & 1  \\
1 & 0 \\
\end{pmatrix}
\ \ \ 
, \sigma_y = 
\begin{pmatrix} 
0 & - i  \\
i & 0 \\
\end{pmatrix}
\ \ \ 
, \sigma_z = 
\begin{pmatrix} 
1 & 0  \\
0 & -1 \\
\end{pmatrix},
\end{equation}
with the $2 \times 2$ identity matrix $\mathbf{I_2}$,
\begin{equation}
\mathbf{M} = \boldsymbol{\sigma} \otimes \mathbf{I_2}  ,  
\ \ \  and \ \ \ M_z = \sigma_z \otimes \mathbf{I_2}.
\end{equation}
The $4 \times 4$ matrices $\boldsymbol{\alpha}$ and $\boldsymbol{\Sigma}$
are given by,
\begin{equation}
\boldsymbol{\alpha} = 
\begin{pmatrix} 
0 & \boldsymbol{\sigma}  \\
\boldsymbol{\sigma} & 0 \\
\end{pmatrix},
\ \ \ 
\boldsymbol{\Sigma} = 
\begin{pmatrix} 
\boldsymbol{\sigma} & 0  \\
0 & \boldsymbol{\sigma} \\
\end{pmatrix}.
\end{equation}
The current density and charge density source matrix is,
\begin{equation}
{W^{\pm} }
=  \frac{1}{\sqrt 2 \epsilon}
\renewcommand\arraystretch{1.6}
\begin{pmatrix}
-J_x \pm i J_y    \\
J_z - v \rho \\
J_z + v \rho \\
J_x \pm i J_y  
\end{pmatrix}
.
\end{equation}

Moreover we shall find that the QLA representation can be determined in modular form :  one need only examine 1D pulse propagation in
each of the 3 orthogonal Cartesian directions.  We explicitly write down these modular components:

\subsection{1D Pulse Propagation in the x-direction}
\begin{equation}
\label{R_S InHomox1}
\frac{\partial}{\partial t}
\begin{bmatrix}
q_0   \\
q_1  \\
q_2   \\
q_3   \\
\end{bmatrix}  	
= - \frac{1}{n(x)} 
\frac{\partial}{\partial x}
\begin{bmatrix}
q_2  \\
q_3  \\
q_0  \\
q_1  \\  
\end{bmatrix}
+  \frac{n^\prime (x)}{2n^2(x)}
\begin{bmatrix}
q_1 + q_6  \\
q_0 - q_7  \\
q_3 - q_4  \\
q_2 + q_5  \\  
\end{bmatrix},
\end{equation}
\begin{equation}
\label{R_S InHomox2}
\frac{\partial}{\partial t}
\begin{bmatrix}
q_4   \\
q_5  \\
q_6   \\
q_7   \\
\end{bmatrix}  	
= - \frac{1}{n(x)} 
\frac{\partial}{\partial x}
\begin{bmatrix}
q_6  \\
q_7  \\
 q_4  \\
 q_5  \\  
\end{bmatrix}
+  \frac{n^\prime (x)}{2n^2(x)}
\begin{bmatrix}
q_5 + q_2  \\
 q_4 - q_3  \\
q_7 - q_0  \\
 q_6 + q_1  \\  
\end{bmatrix}.
\end{equation}

\subsection{1D Pulse Propagation in the y-direction}
\begin{equation}
\label{R_S InHomo1}
\frac{\partial}{\partial t}
\begin{bmatrix}
q_0   \\
q_1  \\
q_2   \\
q_3   \\
\end{bmatrix}  	
= i \frac{1}{n(y)} 
\frac{\partial}{\partial y}
\begin{bmatrix}
q_2  \\
q_3  \\
-q_0  \\
-q_1  \\  
\end{bmatrix}
+ i \frac{n^\prime (y)}{2n^2(y)}
\begin{bmatrix}
q_1 - q_6  \\
-q_0 - q_7  \\
q_3 + q_4  \\
-q_2 + q_5  \\  
\end{bmatrix},
\end{equation}
\begin{equation}
\label{R_S InHomo2}
\frac{\partial}{\partial t}
\begin{bmatrix}
q_4   \\
q_5  \\
q_6   \\
q_7   \\
\end{bmatrix}  	
= i  \frac{1}{n(y)} 
\frac{\partial}{\partial y}
\begin{bmatrix}
-q_6  \\
-q_7  \\
 q_4  \\
 q_5  \\  
\end{bmatrix}
+ i \frac{n^\prime (y)}{2n^2(y)}
\begin{bmatrix}
-q_5 - q_2  \\
 q_4 - q_3  \\
-q_7 + q_0  \\
 q_6 + q_1  \\  
\end{bmatrix}.
\end{equation}

\subsection{1D Pulse Propagation in the z-direction}
\begin{equation}
\label{R_S InHomoz1}
\frac{\partial}{\partial t}
\begin{bmatrix}
q_0   \\
q_1  \\
q_2   \\
q_3   \\
\end{bmatrix}  	
= - \frac{1}{n(z)} 
\frac{\partial}{\partial z}
\begin{bmatrix}
q_0  \\
q_1  \\
-q_2  \\
-q_3  \\  
\end{bmatrix}
+  \frac{n^\prime (z)}{2n^2(z)}
\begin{bmatrix}
q_0 - q_7  \\
-q_1 - q_6  \\
q_2 + q_5  \\
-q_3 + q_4  \\  
\end{bmatrix},
\end{equation}
\begin{equation}
\label{R_S InHomoz2}
\frac{\partial}{\partial t}
\begin{bmatrix}
q_4   \\
q_5  \\
q_6   \\
q_7   \\
\end{bmatrix}  	
= - \frac{1}{n(z)} 
\frac{\partial}{\partial z}
\begin{bmatrix}
q_4 \\
q_5  \\
 -q_6  \\
 -q_7  \\  
\end{bmatrix}
+  \frac{n^\prime (z)}{2n^2(z)}
\begin{bmatrix}
q_4 - q_3  \\
-q_5 - q_2  \\
q_6 + q_1  \\
 -q_7 + q_0  \\  
\end{bmatrix}.
\end{equation}

\section {QLA Representation of Maxwell Equations in Inhomogeneous Media}

From our previous work [3-20] on forming QLA for 1D, 2D, 3D Nonlinear Schrdoinger/Gross Pitaevskii equations, the QLA takes on modular form.
One needs only consider an interleaved sequence of unitary collide-stream operators along a particular axis of the generic form

\begin{equation}
\begin{aligned}
\label{U_X}
 U &= S_{-} C  S_{+} C^{\dagger} \bsdot \: S_{+} C  S_{-} C^{\dagger}, \\
 U^{+} &= S_{+} C^{\dagger}  S_{-}^{01} C \bsdot \: S_{-} C^{\dagger}  S_{+} C
\end{aligned}
\end{equation}
acting on the 8-qubit vector $Q = (q_0 \quad q_1 \quad ...  q_7)^{T}$ to give a time-advancement
\begin{equation}
Q_{t+\delta t} = U^{+} U Q_{t}
\end{equation}
Since all quantum gates can be boiled down to 1-qubit and 2-qubit gates, we need only consider collision operators that couple 2 qubits.
Hence the collision matrices will be, in general, sparse.  The interleaved collide-stream operator sequence is used to recover the differentials in the
partial differential equation of interest in the continuum limit of the lattice equations.  The non-differential terms (in the case of Maxwell equations these are the inhomogeneous refractive index terms) will be modeled by potential matrices $V_{pot}$.  We now turn to specific details for the 3 orthogonal propagation directions.

\subsection{QLA for 1D inhomogeneous Maxwell Equations for x-propagation}
We first consider the construction of the collide-stream operators to recover any partial spatial x-derivatives to second order.  For an 8-qubit vector, the collide-stream matrices will be $8 \cross 8$.

From Eqs. (13) - (14) we see the pairwise coupling of qubits $(q_0 - q_2), (q_1 - q_3), (q_4 - q_6),  (q_5 - q_7) $.   The simplest unitary matrix $C$ with this structure has the form
\begin{equation}
\label{COLL_X 8x8}
C (\theta) = 
  \begin{bmatrix}
  \cos \theta & 0 &  \sin \theta & 0 & 0 & 0 & 0 & 0   \\
  0  &  \cos \theta  &  0  &  \sin \theta  &  0  &  0  &  0  &  0 \\
  - \sin \theta  &  0  &  \cos \theta  &  0  & 0 & 0 & 0 & 0   \\
  0  &  - \sin \theta  &  0  &  \cos \theta  &  0  &  0  &  0  &  0 \\ 
   0 & 0 & 0 & 0  &  \cos \theta & 0 &  \sin \theta & 0 \\
   0 & 0 & 0 & 0  &  0  &  \cos \theta  &  0  &  \sin \theta \\
    0 & 0 & 0 & 0  & - \sin \theta  &  0  &  \cos \theta  &  0 \\
     0 & 0 & 0 & 0  &  0  & - \sin \theta  &  0  &  \cos \theta  \\
   \end{bmatrix}.
\end{equation} 
for some angle $\theta$.   The streaming operator S will stream 4 qubits at a time, and leave the other 4 qubits untouched at their lattice site.  With the specific 2-qubit coupling $(q_0 - q_2), (q_1 - q_3), (q_4 - q_6),  (q_5 - q_7) $  one will choose to stream the 4-qubits $(q_0 \quad q_1 \quad q_4 \quad q_5)$ at one instant and then the other qubits $(q_2 \quad q_3 \quad q_6 \quad q_7)$ at the next instant.  Thus we have 2 basic unitary streaming operators which are diagonal matrices
\begin{equation}
\label{Streamx012}
S_{\pm 1}^{0145}
\begin{bmatrix}
q_0 (x)   \\
q_1  (x)\\
q_2  (x) \\
q_3  (x)  \\
q_4  (x) \\
q_5  (x)\\
q_6  (x)\\
q_7  (x) \\
\end{bmatrix}  	
= 
 \begin{bmatrix}
q_0 (x \pm 1)   \\
q_1  (x \pm 1)\\
q_2  (x) \\
q_3  (x)  \\
q_4  (x \pm 1) \\
q_5  (x \pm 1)\\
q_6  (x)\\
q_7  (x) \\
\end{bmatrix},
\quad
S_{\pm 1}^{2367}
\begin{bmatrix}
q_0 (x)   \\
q_1  (x)\\
q_2  (x) \\
q_3  (x)  \\
q_4  (x) \\
q_5  (x)\\
q_6  (x)\\
q_7  (x) \\
\end{bmatrix}  	
= 
 \begin{bmatrix}
q_0 (x)   \\
q_1  (x)\\
q_2  (x  \pm 1) \\
q_3  (x  \pm 1)  \\
q_4  (x) \\
q_5  (x)\\
q_6  (x  \pm 1)\\
q_7  (x  \pm 1) \\
\end{bmatrix}
\end{equation}

If we now define the unitary interleaved sequence of collision-stream operators
\begin{equation}
\begin{aligned}
  U = S_{- \epsilon} ^{0145} . C(\theta) .  S_{+ \epsilon} ^{0145} . C^{+}(\theta) .  S_{+ \epsilon} ^{2367} . C(\theta) .  S_{- \epsilon} ^{2367} . C^{+}(\theta) ,  \\
   U^{+} = S_{+\epsilon} ^{0145} . C^{+}(\theta) .  S_{- \epsilon} ^{0145} . C(\theta) .  S_{- \epsilon} ^{2367} . C^{+}(\theta) .  S_{+ \epsilon} ^{2367} . C(\theta) 
\end{aligned}
\end{equation}
and symbolically evaluate
\begin{equation}
  U^{+} . U . Q(x,t)
\end{equation}
for what we can define as one time instant $\delta t$ of propagation, we find
\begin{equation}
\label{Algorx}
\begin{bmatrix}
q_0 (x,t+\delta t)   \\
q_1  (x,t+\delta t)\\
q_2  (x,t+\delta t) \\
q_3  (x,t+\delta t)  \\
q_4  (x,t+\delta t) \\
q_5  (x,t+\delta t)\\
q_6  (x,t+\delta t)\\
q_7  (x,t+\delta t) \\
\end{bmatrix}  	
= 
 \begin{bmatrix}
q_0 (x,t)   \\
q_1  (x,t)\\
q_2  (x,t) \\
q_3  (x,t)  \\
q_4  (x,t) \\
q_5  (x,t)\\
q_6  (x,t)\\
q_7  (x,t) \\
\end{bmatrix}
-
\frac{1}{n(x)}    \frac{\partial}{\partial x}
 \begin{bmatrix}
q_2 (x,t)   \\
q_3  (x,t)\\
q_0  (x,t) \\
q_1  (x,t)  \\
q_6  (x,t) \\
q_7  (x,t)\\
q_4  (x,t)\\
q_5  (x,t) \\
\end{bmatrix}
\epsilon^2 + O(\epsilon^4)
\end{equation}
on choosing the collision angle 
\begin{equation}
\theta = \frac{\epsilon}{4 n(x)} , \quad  with  \quad  \epsilon << 1.
\end{equation}
Thus to recover the required differentials in the long-time long-wavelength continuum limit we must enforce
diffusion ordering on the time scales  $\delta t = \epsilon^2$, where the spatial lattice unit spacing is $\epsilon$ :
\begin{equation}
\label{R_S InHomoxxx1}
\frac{\partial}{\partial t}
\begin{bmatrix}
q_0(x,t)   \\
q_1(x,t)   \\
q_2(x,t)    \\
q_3 (x,t)   \\
q_4  (x,t)  \\
q_5 (x,t)  \\
q_6 (x,t)   \\
q_7 (x,t)   \\
\end{bmatrix}  	
= - \frac{1}{n(x)} 
\frac{\partial}{\partial x}
\begin{bmatrix}
q_2 (x,t)   \\
q_3 (x,t)  \\
q_0 (x,t)  \\
q_1 (x,t)  \\ 
q_6 (x,t)   \\
q_7 (x,t)  \\
q_4 (x,t)  \\
q_5 (x,t)  \\  
\end{bmatrix}
+ O(\epsilon^2) .
\end{equation}
So far the QLA is unitary.

To recover the two terms associated with the inhomogeneous refractive index
\begin{equation}
 \frac{n^{\prime}(x)}{2 n^2 (x)}
\end{equation}
we will introduce two potential-like collision operators.  Their basic form can be deduced from the required couplings in the
x-dependent Maxwell equations.  The first potential collision operator couples qubits $(q_0 - q_1, q_2 - q_3, q_4 - q_5, q_6 - q_7)$
while the second potential collision operator couples qubits $(q_0 - q_6, q_1 - q_7 , q_2 - q_4, q_3 - q_5)$.  An appropriate choice for these
$8 \cross 8$ matrices is
\begin{equation}
\label{COLL_Y 8x8}
V_1 (\alpha) = 
  \begin{bmatrix}
  \cos \alpha & -\sin \alpha &  0 & 0 & 0 & 0 & 0 & 0   \\
  -\sin \alpha  &  \cos \alpha  &  0  &  0  &  0  &  0  &  0  &  0 \\
 0 &  0  &  \cos \alpha   &   -\sin \alpha   & 0 & 0 & 0 & 0   \\
  0  &  0 &   -\sin \alpha   &  \cos \alpha  &  0  &  0  &  0  &  0 \\ 
   0 & 0 & 0 & 0  &  \cos \alpha &  -\sin \alpha  &  0 & 0 \\
   0 & 0 & 0 & 0  &   -\sin \alpha   &  \cos \alpha  &  0  &  0 \\
    0 & 0 & 0 & 0  & 0 &  0  &  \cos \alpha  &   -\sin \alpha  \\
     0 & 0 & 0 & 0  &  0  & 0 &   -\sin \alpha   &  \cos \alpha  \\
   \end{bmatrix}.
\end{equation} 
and
\begin{equation}
V_2 (\alpha) = 
  \begin{bmatrix}
  \cos \alpha & 0 &  0 & 0 & 0 & 0 & -\sin \alpha  & 0   \\
  0 &  \cos \alpha  &  0  &  0  &  0  &  0  &   0   & \sin \alpha  \\
 0 &  0  &  \cos \alpha   &  0 &  \sin \alpha  & 0 & 0 & 0    \\
  0  &  0 & 0   &  \cos \alpha  &  0  &   -\sin \alpha   &  0  &  0 \\ 
   0 & 0 &  -\sin \alpha  & 0  &  \cos \alpha & 0 &  0 & 0 \\
   0 & 0 & 0 &  \sin \alpha   &   0   &  \cos \alpha  &  0  &  0 \\
     \sin \alpha  & 0 & 0 & 0  & 0 &  0  &  \cos \alpha  & 0  \\
     0 &  -\sin \alpha & 0 & 0  &  0  & 0 &  0 &  \cos \alpha  \\
   \end{bmatrix}.
\end{equation} 
for some angle $\alpha$.  From symbolic manipulations one finds that the appropriate potential collision angle $\alpha$ is
\begin{equation}
 \alpha = \epsilon^2 \frac{n^{\prime}(x)}{2 n^2 (x)}. 
\end{equation}

It is interesting to note that while $V_2(\alpha)$ is unitary, the potential collision martix $V_1(\alpha)$ is not unitary, but just Hermitian.

The final QLA for 1D propagation in an x-dependent inhomogeneous medium is thus
\begin{equation}
Q(t + \delta t) = V_2(\alpha). V_1(\alpha). U^{+}(\theta). U(\theta). Q(t)
\end{equation}

\subsection{QLA for 1D inhomogeneous Maxwell Equations for y-propagation}

As can be seen from Eqs. (15)-(16), the y-dependent 1D Maxwell equations are very similar to those for the x-dependent ones.  Of course, this is predicated by the similarity of the Pauli spin matrices $\sigma_x$ and $\sigma_y$ :
\begin{equation}
\label{Pauli_spin}
\sigma_x = 
\begin{pmatrix} 
0 & 1  \\
1 & 0 \\
\end{pmatrix}
\ \ \ 
, \sigma_y = 
\begin{pmatrix} 
0 & - i  \\
i & 0 \\
\end{pmatrix}
\end{equation}

Hence we will simply write down the corresponding unitary collision-streaming operators.  
\begin{equation}
\label{COLL_Y 8x8}
C (\theta) = 
  \begin{bmatrix}
  \cos \theta & 0 & i \sin \theta & 0 & 0 & 0 & 0 & 0   \\
  0  &  \cos \theta  &  0  & i \sin \theta  &  0  &  0  &  0  &  0 \\
  i \sin \theta  &  0  &  \cos \theta  &  0  & 0 & 0 & 0 & 0   \\
  0  &  i \sin \theta  &  0  &  \cos \theta  &  0  &  0  &  0  &  0 \\ 
   0 & 0 & 0 & 0  &  \cos \theta & 0 & -i \sin \theta & 0 \\
   0 & 0 & 0 & 0  &  0  &  \cos \theta  &  0  &  -i\sin \theta \\
    0 & 0 & 0 & 0  & -i \sin \theta  &  0  &  \cos \theta  &  0 \\
     0 & 0 & 0 & 0  &  0  & -i \sin \theta  &  0  &  \cos \theta  \\
   \end{bmatrix}.
\end{equation} 
Since the collision operator couples the same two qubits as for the x-dependent Maxwell equation, the streaming operators will be unchanged.
Moreover, for the inhomogeneous refractive index the two qubit coupings are also unchanged.  Hence the two potential collision matrices are
\begin{equation}
\label{V11 8x8}
V_{1} (\beta) = 
\begin{bmatrix}
\cos \beta & \sin \beta & 0 & 0 & 0 & 0 & 0 & 0   \\
-\sin \beta  &  \cos \beta  &  0  & 0  &  0  &  0  &  0  &  0 \\
0  &  0  &  \cos \beta & \sin \beta & 0 & 0 & 0 & 0   \\
0  &  0  &  -\sin \beta & \cos \beta & 0 & 0 & 0 & 0  \\ 
0 & 0 & 0 & 0  &  \cos \beta & -\sin \beta & 0 & 0 \\
0 & 0 & 0 & 0  &  \sin \beta  &  \cos \beta  &  0  & 0 \\
0 & 0 & 0 & 0  & 0  &  0  &  \cos \beta & -\sin \beta \\
0 & 0 & 0 & 0  &  0  & 0  &  \sin \beta & \cos \beta  \\
\end{bmatrix},
\end{equation} 
\begin{equation}
\label{V22 8x8}
V_{2} (\beta) = 
\begin{bmatrix}
\cos \beta & 0 & 0 & 0 & 0 & 0 & -\sin \beta & 0   \\
0  &  \cos \beta  &  0  & 0  &  0  &  0  &  0  &  -\sin \beta \\
0  &  0  &  \cos \beta & 0 & \sin \beta & 0 & 0 & 0   \\
0  &  0  &  0 & \cos \beta & 0 & \sin \beta & 0 & 0  \\ 
0 & 0 & -\sin \beta & 0  &  \cos \beta & 0 & 0 & 0 \\
0 & 0 & 0 & -\sin \beta  &  0  &  \cos \beta  &  0  & 0 \\
\sin \beta & 0 & 0 & 0  & 0  &  0  &  \cos \beta & 0 \\
0 & \sin \beta & 0 & 0  &  0  & 0  & 0 & \cos \beta  \\
\end{bmatrix}.
\end{equation}
The 1D y-dependent Maxwell equations are recovered from this QLA provided the collision angles
\begin{equation}
\theta = \frac{\epsilon}{4 n(y)} ,  \quad   \beta = - i \epsilon^2 \frac{n^{\prime} (y)}{2n^2(y)}.
\end{equation}
Because of the complex collision angle $\beta$, both the potential collision matrices are Hermitian, but not unitary.  The final QLA for y-dependent refractive index has a slightly different collide-stream interleaved sequence
\begin{equation}
\begin{aligned}
  U = S_{- \epsilon} ^{2367} . C(\theta) .  S_{+ \epsilon} ^{2367} . C^{+}(\theta) .  S_{+ \epsilon} ^{0145} . C(\theta) .  S_{- \epsilon} ^{0145} . C^{+}(\theta) ,  \\
   U^{+} = S_{+\epsilon} ^{2367} . C^{+}(\theta) .  S_{- \epsilon} ^{2367} . C(\theta) .  S_{- \epsilon} ^{0145} . C^{+}(\theta) .  S_{+ \epsilon} ^{0145} . C(\theta) 
\end{aligned}
\end{equation}
so that for the 8-qubit vector Q:
\begin{equation}
Q(t + \delta t) = V_2(\beta). V_1(\beta). U^{+}(\theta). U(\theta). Q(t)
\end{equation}

\subsection{QLA for 1D inhomogeneous Maxwell Equations for z-propagation}

The QLA for z-dependent propagation is very different from that for the other two orthogonal directions.  This is because the Pauli spin matrix $\sigma_z$ is diagonal
\begin{equation}
\label{Pauli_spin z}
\sigma_z = 
\begin{pmatrix} 
1 & 0  \\
0 & -1 \\
\end{pmatrix}
\end{equation} 
resulting in the coupling of $\partial q_i / \partial t$  to  $\partial q_i / \partial z$, for each i, i = 0 ... 7.
Since the unitary collision operators must couple two different qubits, there is no $8 \cross 8$ representation for z-propagation.  Hence we turn to a
16-qubit representation.  In our earlier work on developing QLA for solitons and Bose-Einstein condensates [15-16], the physical order-parameter equations (nonlinear Schrodinger equation or the Gross-Pitaevskii equation) were represented at a mesoscopic level by twice as many qubits as field components.  Especially when dealing with a single scalar field equation, one needed at least 2 qubits per spatial lattice grid to represent the field so that there could be quantum entanglement arising from the unitary qubit collision operator.  Because of vector nature of the electromagnetic fields in Maxwell equations there were sufficient qubits per lattice site to represent the fields directly.  However, for z-dependent propagation, the diagonal form of the Pauli spin matrix $\sigma_z$ forces us into a mesoscopic qubit representation.

An appropriate unitary collision matrix which couples qubits $(q_0 - q_2, q_1 - q_3,  q_5 - q_7, q_6 - q_8, q_9 -q_{11}, q_{12} - q_{14}, q_{13} - q_{15})$ has the following
$4 \cross 4$ block structure
\begin{equation}
\label{C4x4}
C (\theta) = 
\begin{bmatrix}
V_4 (\theta) & 0 & 0 & 0   \\
0  &  V_4 (\theta)^{Tr}  & 0 & 0  \\  
0  & 0 & V_4 (\theta)  &  0 \\
 0  &  0 & 0 & V_4 (\theta)^{Tr} \\ 
\end{bmatrix},
\end{equation} 
where
\begin{equation}
\label{Csub4x4}
V_4 (\theta) = 
\begin{bmatrix}
\cos \theta & 0 & \sin \theta & 0   \\
0  &  \cos \theta  & 0 & \sin \theta  \\  
-\sin \theta  & 0 &  \cos \theta  &  0 \\
 0  &  -\sin \theta & 0 & \cos \theta  \\ 
\end{bmatrix},
\end{equation} 
and $V_4 (\theta)^{Tr}$ is the transpose of the $4 \cross 4$ matrix $V_4 (\theta)$, Eq. (42).

The streaming operators each stream 8 qubits : let us denote one of these operators $S^{08}$ which streams qubits $(q_0, q_1, q_4, q_5, q_8, q_9, q_{12}, q_{13})$  while $S^{2 \:10}$ streams the 8 qubits $(q_2, q_3, q_6,q_7, q_{10}, q_{11}, q_{14},q_{15})$.  The first of the potential collision operators mimics the coupling of the unitary collision matrix $C$ so its $4 \cross 4$ block structure is 
\begin{equation}
\label{C4x4}
P_1 (\gamma) = 
\begin{bmatrix}
PV_4 (\gamma) & 0 & 0 & 0   \\
0  &  PV_4 (\gamma) & 0 & 0  \\  
0  & 0 & PV_4 (\gamma) &  0 \\
 0  &  0 & 0 & PV_4 (\gamma) \\ 
\end{bmatrix},
\end{equation} 
where
\begin{equation}
\label{Csub4x4}
PV_4 (\gamma) = 
\begin{bmatrix}
\cos \gamma & 0 &- \sin \gamma & 0   \\
0  &  \cos \theta  & 0 & - \sin \gamma \\  
- \sin \gamma & 0 & \cos \gamma  &  0 \\
 0  &  - \sin \gamma & 0 &\cos \gamma \\ 
\end{bmatrix},
\end{equation}
while the second  potential collision operator has diagonal-like structure of two $8 \cross 8$ matrices
\begin{equation}
\label{Csub4x4}
P_2 (\gamma) = 
\begin{bmatrix}
PV_{81} (\gamma) & PV_{82} (\gamma)   \\
PV_{82} (\gamma) & PV_{81} (\gamma)  \\  
\end{bmatrix},
\end{equation}
where
\begin{equation}
\label{PV81 8x8}
PV_{81} (\gamma) = 
  \begin{bmatrix}
  \cos \gamma & 0 & 0 & 0 & 0 & 0 & 0 & 0   \\
  0  &  \cos \gamma  &  0  & 0  &  0  &  0  &  0  &  0 \\
  0  &  0  &  \cos \gamma  &  0  & 0 & 0 & 0 & 0   \\
  0  &  0  &  0  &  \cos \gamma  &  0  &  0  &  0  &  0 \\ 
   0 & 0 & 0 & 0  &  \cos \gamma & 0 &  0 & 0 \\
   0 & 0 & 0 & 0  &  0  &  \cos \gamma &  0  &  0 \\
    0 & 0 & 0 & 0  & 0  &  0  &  \cos \gamma  &  0 \\
     0 & 0 & 0 & 0  &  0  & 0  &  0  &  \cos\gamma  \\
   \end{bmatrix}.
\end{equation} 
and
\begin{equation}
\label{PV82 8x8}
PV_{82} (\gamma) = 
  \begin{bmatrix}
 0 & 0 & 0 & 0 & 0 & 0 & 0 & - \sin \gamma   \\
  0  & 0  &  0  & 0  &  0  &  0  &  - \sin \gamma  &  0 \\
  0  &  0  & 0 &  0  & 0 & - \sin \gamma & 0 & 0   \\
  0  &  0  &  0  & 0 &  - \sin \gamma  &  0  &  0  &  0 \\ 
   0 & 0 & 0 &  \sin \gamma   &  0 & 0 &  0 & 0 \\
   0 & 0 &  \sin \gamma  & 0  &  0  &  0 &  0  &  0 \\
    0 &  \sin \gamma  & 0 & 0  & 0  &  0  &  0  &  0 \\
      \sin \gamma  & 0 & 0 & 0  &  0  & 0  &  0  &  0  \\
   \end{bmatrix}.
\end{equation} 

With the unitary operators
\begin{equation}
\begin{aligned}
  U_{[16]} =S_{-\epsilon}^{0,8}.C(\theta). S_{+\epsilon}^{0,8} . C^{+}(\theta).S_{+\epsilon}^{2,10}.C(\theta). S_{-\epsilon}^{2,10} . C^{+}(\theta)  \\
  U_{[16]}^{+} =  S_{+\epsilon}^{0,8}.C^{+}(\theta). S_{-\epsilon}^{0,8} . C(\theta).S_{-\epsilon}^{2,10}.C^{+}(\theta). S_{+\epsilon}^{2,10} . C(\theta)
\end{aligned}
\end{equation}
one obtains from
\begin{equation}
   Q_{[16]}(t + \delta t) = P_2(\gamma) P_1(\gamma) U_{]16]}^{+}. U_{[16]} . Q_{[16]} (t)
\end{equation}
on using the collision angles
\begin{equation}
 \theta = \frac{\epsilon}{4 n(z)} ,  \quad \gamma = \frac{\epsilon^2 n^\prime{z}}{2 n^2{z}}
\end{equation}
the mesoscopic evolution of the 16-qubits.  This evolution falls into a 4-block structure of the form
\begin{equation}
\begin{aligned}
  q_{0+4k}(t+ \delta t)  = q_{0+4k}(t) + \Big\{ \frac{n^{\prime}(z)}{4 n^2(z)}  [-q_{2+4k}(z) +(-1)^{k} q_{15-4k}(z) ] - (-1)^{k}\frac{1}{4 n(z)} \frac{\partial q_{2+4k}}{\partial z} \Big \} \epsilon^2 + O(\epsilon^4) \\
    q_{1+4k}(t+ \delta t)  = q_{1+4k}(t) + \Big\{ \frac{n^{\prime}(z)}{4 n^2(z)}  [+q_{3+4k}(z) + (-1)^{k} q_{14-4k}(z) ] - (-1)^{k} \frac{1}{4 n(z)} \frac{\partial q_{3+4k}}{\partial z} \Big \} \epsilon^2 + O(\epsilon^4) \\
  q_{2+4k}(t+ \delta t)  = q_{2+4k}(t) + \Big\{ \frac{n^{\prime}(z)}{4 n^2(z)}  [-q_{0+4k}(z) + (-1)^{k} q_{13-4k}(z) ] -(-1)^{k} \frac{1}{4 n(z)} \frac{\partial q_{0+4k}}{\partial z} \Big \} \epsilon^2 + O(\epsilon^4) \\
  q_{3+4k}(t+ \delta t)  = q_{3+4k}(t) + \Big\{ \frac{n^{\prime}(z)}{4 n^2(z)}  [+q_{1+4k}(z) +(-1)^{k} q_{12-4k}(z) ] -(-1)^{k} \frac{1}{4 n(z)} \frac{\partial q_{1+4k}}{\partial z} \Big \} \epsilon^2 + O(\epsilon^4) \\
\end{aligned}
\end{equation}
with k = 0, 1, 2 or 3.

To recover the 8-spinor representation of the 1D Maxwell equations in Riemann-Silberstein form we need only define 

\begin{equation}
\begin{aligned}
\overline{q_0} = q_0 + q_2 ,\quad  \overline{q_1} = q_1 + q_3.  \quad. \overline{q_2} = q_4 + q_6 , \quad \overline{q_3} = q_5 + q_7. \\
\overline{q_4} = q_8 + q_{10} , \quad  \overline{q_5} = q_9 + q_{11} , \quad \overline{q_6} = q_{12} + q_{14} , \quad \overline{q_7} = q_{13} + q_{15} . \\
\end{aligned}
\end{equation}

\section{Some Simulations for Electromagnetic Pulses propagating in the z-direction}
We have previously considered the QLA for 1D Maxwell equations in inhomogeneous media for propagation in the y-direction [17] and in the x-direction [31].  Both these QLAs require an 8-qubit representation.  Here we present some longer time evolution of z-propagating Gaussian pulse which now requires 16 qubits/node. In particular we shall consider multiple reflections and transmissions at strongly varying dielectric boundary layers.  We shall consider an inhomogeneous medium with vacuum refractive index $n(z) = 1.0 $ for $0 < z < 3700$ and for $4300 < z < 6500$ and  refractive index $n(z) = 2 $ for $3700 < z < 4300$, Fig. 1.

\begin{figure}[!h!p!b!t] \ 
\begin{center}
\includegraphics[width=5.1in]{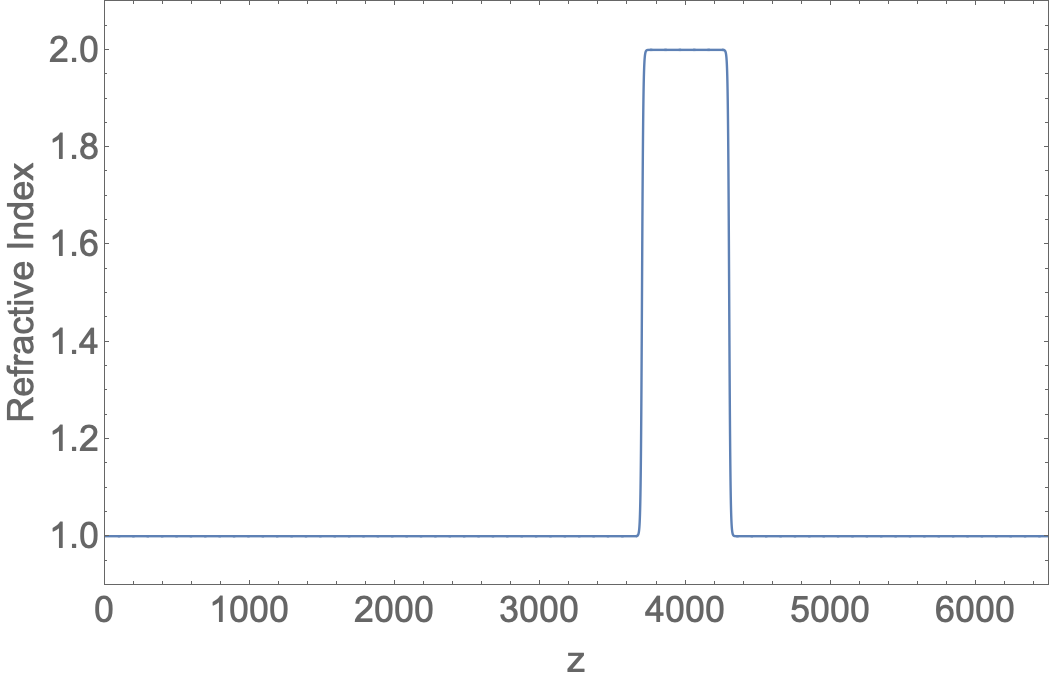}
	\caption{A localized inhomogeneous dielectric region, $3700 < z < 4300$ within a vacuum.
}
\end{center}
\end{figure}

\subsection{Normal incident electromagnetic pulse}
The simplest Gaussian vacuum electromagnetic pulse propagating in the z-direction has for the non-zero components of the electric $\mathbf{E}$ and magnetic $\mathbf{B}$ fields
\begin{equation}
E_x(z,t=0) = 0.01 \;exp[- \frac{\epsilon^2 (z-z_0)^2}{1500}] = B_y(z,t=0)
\end{equation}
where the small parameter $\epsilon = 0.3$ and the initial center of the Gaussian pulse is at $z_0 = 2300$.

This incident pulse propagates undistorted towards the dielectric slab, as seen in Fig. 2 for times $t = 0$, and $t < 4000$. By $t = 4000$ the forward part of the pulse is just starting to interact with the dielectric boundary layer.
\begin{figure}[!h!p!b!t] \ 
\begin{center}
\includegraphics[width=5.1in]{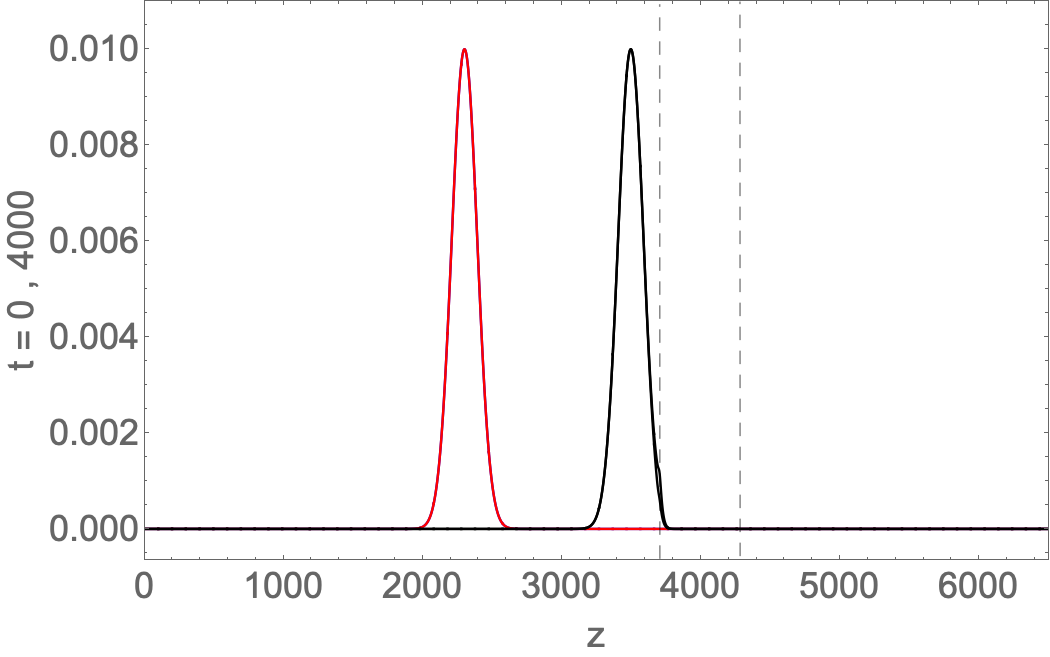}
	\caption{ The Gausssian pulse at $t = 0 (red)$ and $t = 4000 (black)$.  In the vacuum, $E_x = B_y$ and these fields overlay each other.  The vertical dashed lines indicate the dielectric slab of refractive index $n = 2$.
}
\end{center}
\end{figure}
Fig. 3 shows the pulse straddling the vacuum-dielectric slab region at time $t = 4800$.  Within the dielectric slab, the transmitted pulse has  $E_x \ne B_y$, with $max \, B_y = 2 \, max \,E_x$.  The part of the pulse in the vacuum is predominantly the transient reflected part with the beginnings of a phase shift in $E_x$ since the reflection is occurring at a low-to-higher refractive index.  
\begin{figure}[!h!p!b!t] \ 
\begin{center}
\includegraphics[width=5.1in]{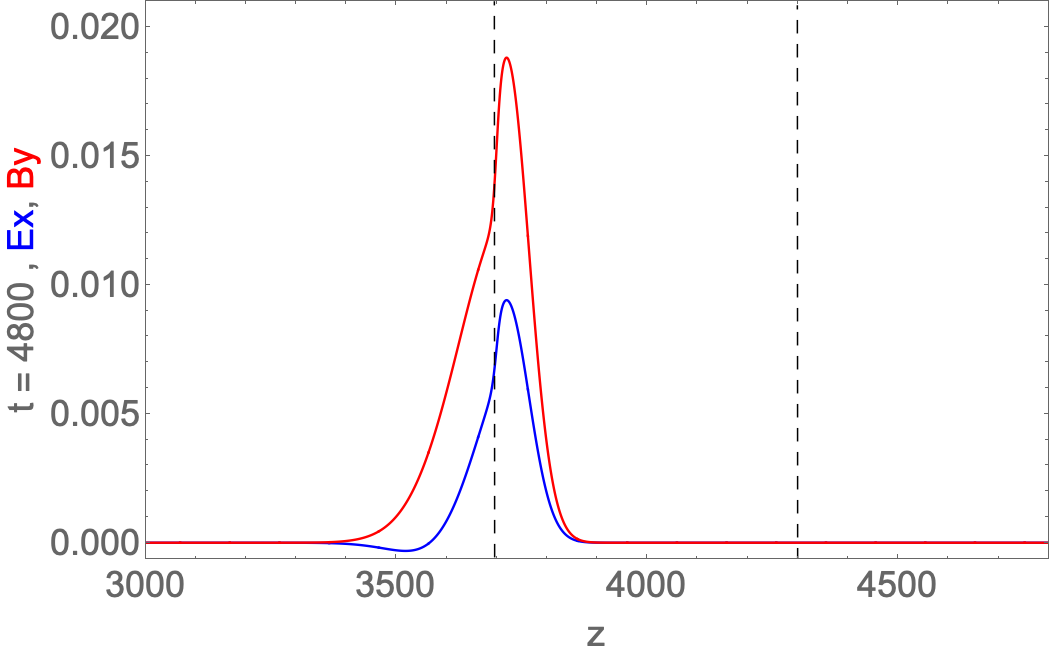}
\caption{ At t = 4800, the incident vacuum pulse is being split into a partly transmitted pulse within the dielectric slab of $n_{slab} = 2$, and partially reflected pulse back into the vacuum.  Because the pulse is moving from vacuum into a higher refractive index region it is the electric field component $E_x$ that exhibits phase change.  ($E_x - blue, B_y - red$ , dielectric slab lies in $3700 < z < 4300$, its boundaries denoted by the dashed vertical lines).
}
\end{center}
\end{figure}
By $t = 6000$ the transients have died down and one sees the transmitted pulse within the dielectric slab, with $B_y \simeq E_x$ and in phase, while the reflected pulse back into the vacuum has $E_x \pi$ out of phase with $B-y$.  The speed of the transmitted pulse is half that of the incident and reflected pulse in the vacuum (Fig. 4).
\begin{figure}[!h!p!b!t] \ 
\begin{center}
\includegraphics[width=5.1in]{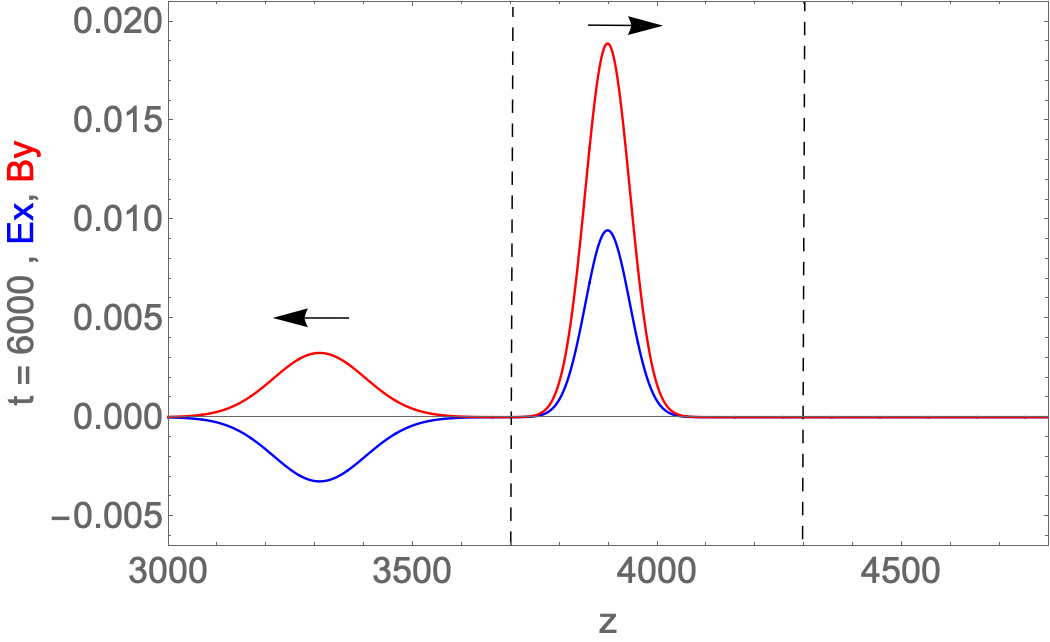}
\caption{The reflected and transmitted pulses at time $t = 6000$.  Since the initial pulse is incident onto a higher refractive index medium, the reflected $E_x$ undergoes a $\pi$ phase change, but with $\abs{E^{refl}_x} = B^{ref;}_y$.  The transmitted pulse's speed is half of the initial vacuum pulse's, and half the initial pulse width,  with $B^{trans}_y \simeq 2  E^{trans}_x$.  ($E_x - blue, B_y - red$ , dielectric slab lies in $3700 < z < 4300$, its boundaries denoted by the dashed vertical lines).
}
\end{center}
\end{figure}

By $t = 9000$ the transmitted pulse has reached the right boundary of the dielectric slab and it is undergoing its transmission and reflection, Fig. 5.  Since the reflected pulse is going back into a higher refractive medium it is now $B_y$ that undergoes a $\pi$ phase change.
\begin{figure}[!h!p!b!t] \ 
\begin{center}
\includegraphics[width=5.1in]{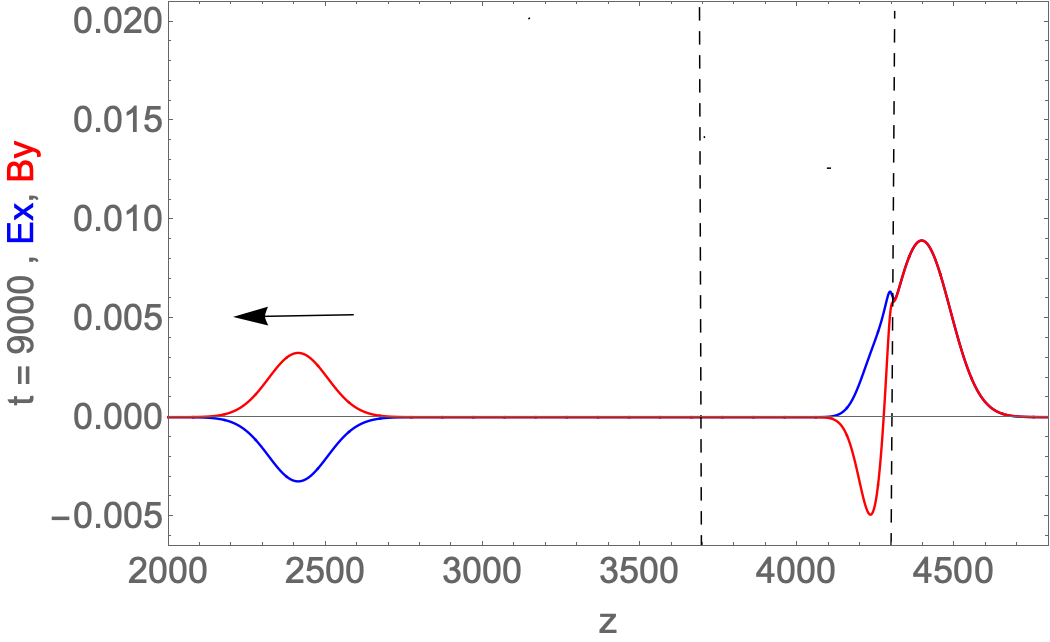}
\caption{By t = 9000, the right traveling pulse reaches  the right end of the dielectric slab and it itself undergoes transient transmission and reflection.  Now, however it is the magnetic field $B_y$ that undergoes a  phase change  ($E_x - blue, B_y - red$ , dielectric slab lies in $3700 < z < 4300$, its boundaries denoted by the dashed vertical lines).
}
\end{center}
\end{figure}

The time asymptotic state of this stage of pulse evolution is shown in Fig. 6 ($t = 10000$).  The transmitted pulse ($z > 4300$ is back in the vacuum region so that $E_x \simeq B_y$ and the field components overlay each other, as at t = 0.  This pulse has the same speed and width of the initial vacuum pulse, but its amplitude is somewhat reduced (to preserve total energy conservation).  The reflected pulse in the dielectric slab is propagating to the left with half the speed of the initial vacuum pulse, with half its width and now with the $B_y$ out-of-phase by $\pi$ with its $E_x$ field component.
\begin{figure}[!h!p!b!t] \ 
\begin{center}
\includegraphics[width=5.1in]{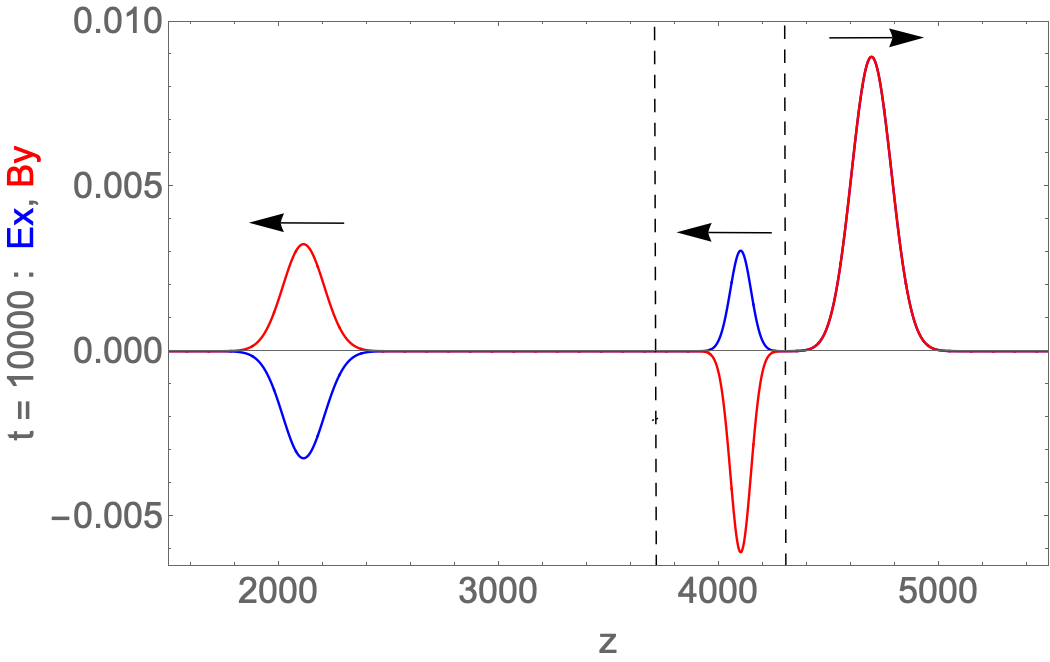}
\caption{A quasi-asymptotic state ($t = 10000$) with the first reflection and transmission off the back interface around $z=4300$.  For the reflect pulse  in the dielectric slab, it is the magnetic field $B_y$ that undergoes a $\pi-$ phase transition.  ($E_x - blue, B_y - red$ , dielectric slab lies in $3700 < z < 4300$, its boundaries denoted by the dashed vertical lines).
}
\end{center}
\end{figure}

In our simulations we then follow the reflection and transmission of the left-traveling pulse within the dielectric as it hits the inner edge around $z=3700$.  The subsequent transmitted pulse then keeps propagating to the left into the vacuum has its $B_y$ out of phase with its companion $E_x$.  The part of the pulse that is reflected from the dielectric boundary around $z=3700$ has another $\pi-$ phase change induced in the magnetic field component $B_y$ so that this right traveling pulse within the dielectric has its components again in phase (see Fig. 7)
 \begin{figure}[!h!p!b!t] \ 
\begin{center}
\includegraphics[width=5.1in]{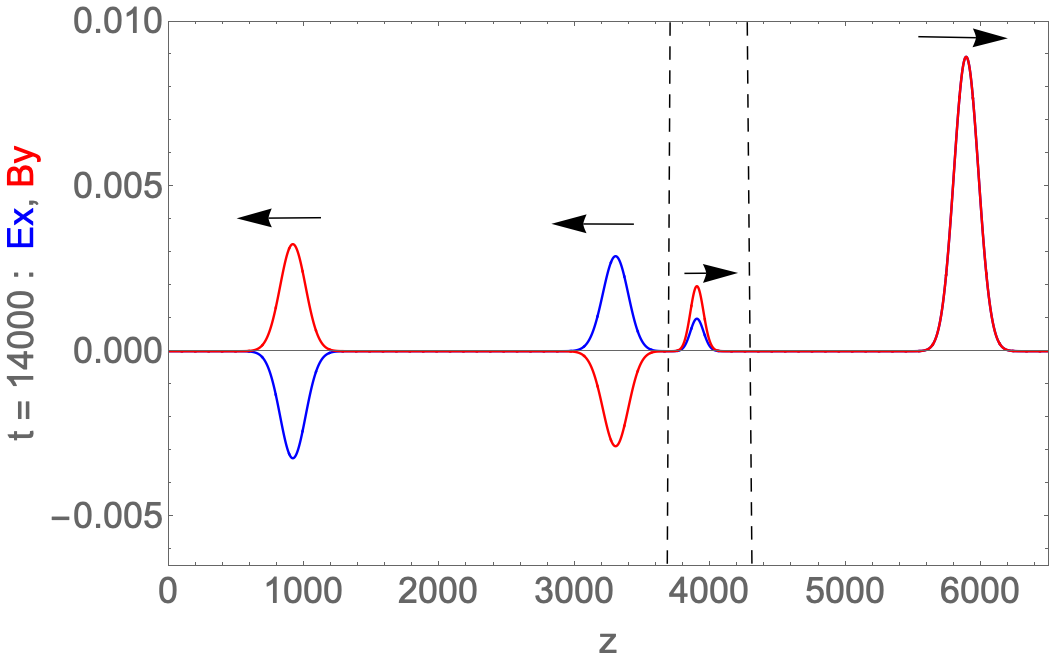}
\caption{A quasi-asymptotic state ($t = 14000$) following a reflection and transmission off the front interface around $z=3700$. In the outgoing pulses traveling to the left, the two pulses are out of phase with each other.  ($E_x - blue, B_y - red$ , dielectric slab lies in $3700 < z < 4300$, its boundaries denoted by the dashed vertical lines).
}
\end{center}
\end{figure}             

Finally, in these 1D simulations of the normal incidence of a Gaussian pulse onto a dielectric slab we computer the instantaneous Poynting flux S(t)
\begin{equation}
S(t)= \int_0^L \mathbf{E}(z,t) \times \mathbf{ B}_y(z,t) \cdot \mathbf{n} \;dz
\end{equation}
It is seen from Fig. 8, that QLA conserves energy very well throughout the simulation except during the overlap of incident/reflected pulses around the slab boundaries $z=3700$ or $z=4300$.  During these time intervals it is very difficult to distinguish which part of the pulse is incident and which part of the pulse is due to reflection -- see e.g., Fig. 3 (for $t=4800$) or FIg. 5 ($t=9000$) - thus making the identification of the outward pointing normal  difficult.
 \begin{figure}[!h!p!b!t] \ 
\begin{center}
\includegraphics[width=5.1in]{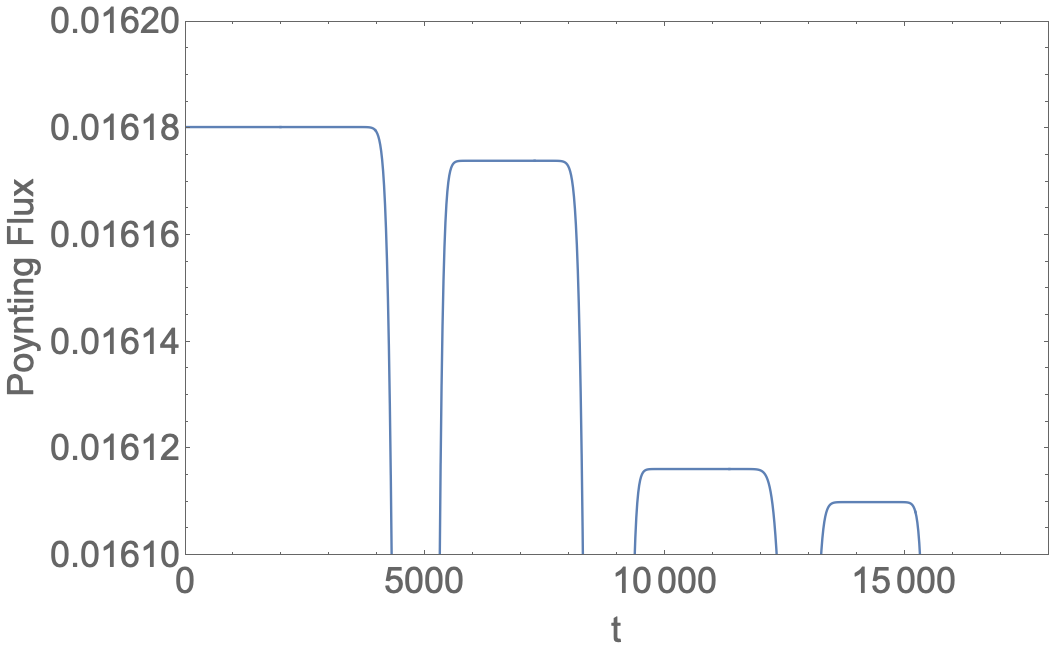}
\caption{The instantaneous Poynting flux.  The time intervals $4200 < t < 5300$, $8200 < t < 9300$ and $12400 < t < 13300$ are those in which there is pulse overlap with either the front or the back boundaries of the dielectric slab.
}
\end{center}
\end{figure}  

\section{Preliminary 2D QLA Scattering from a Dielectric Obstacle}
We can now readily stitch together the various three orthogonal QLAs to obtain a 2D or 3D Maxwell solver for electromagnetic fields in an arbitrary scalar dielectric medium.  Here some preliminary 2D $x-z$ QLA simulations for an initial Gaussian pulse propagating towards a conical dielectric obstacle are presented.  The $x-z$ QLA is obtained simply by stitching together the evolution equations (32) and (49). A 2D dielectric cone is situated with a base $75 < x/4 < 225$ and $175 < z/4 < 325$ and rising to a refractive index of 3 around $x/4 = 150, z/4 = 255$.  The simulations were performed on a $2028 \times 2048$ grid, but the data was plotted at every 4th point - hence the figures are drawn on a $0 < x/4, z/4 < 512$ grid.

\begin{figure}[!h!p!b!t] \ 
\begin{center}
\includegraphics[width=5.1in]{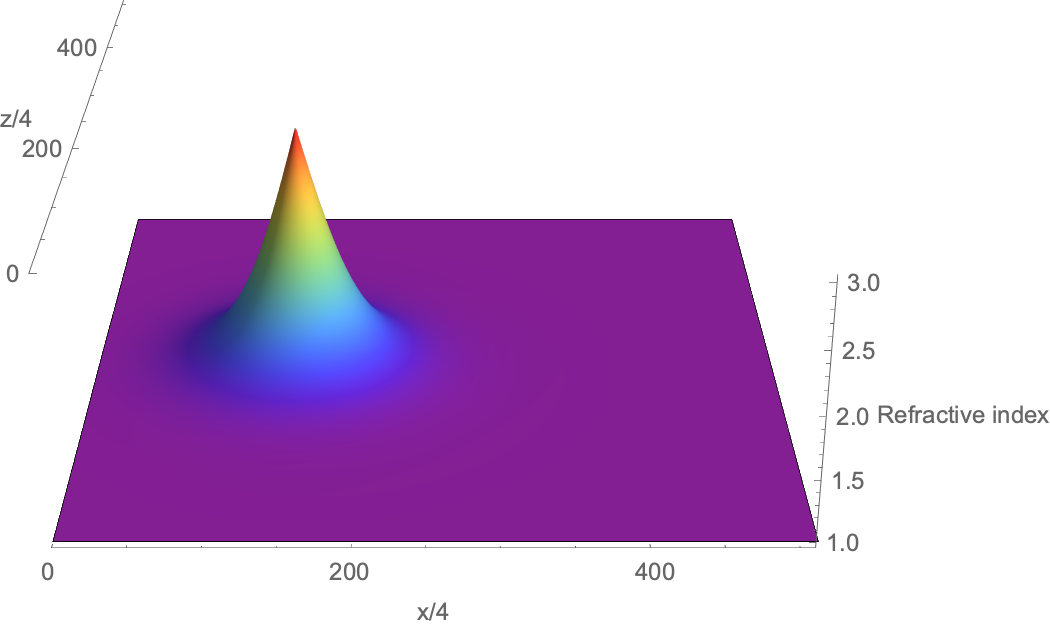}
\caption{The refractive index of a conical dielectric obstacle 
}
\end{center}
\end{figure} 

The initial Gaussian plane pulse is independent of $z$ and propagates in the x-direction towards the dielectric obstacle, Fig. 10.
\begin{figure}[!h!p!b!t] \ 
\begin{center}
\includegraphics[width=5.1in]{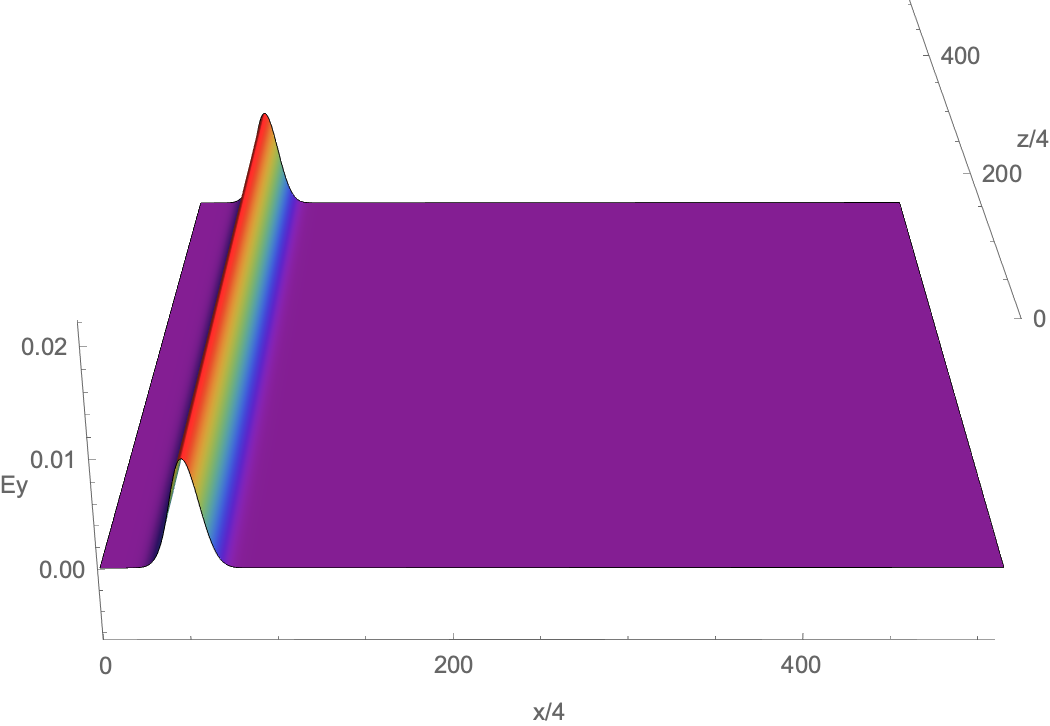}
\caption{The initial Gaussian pulse propagating in the $x-$ direction.
}
\end{center}
\end{figure} 

By $t=1250$, the Gaussian pulse is interacting with the dielectric cone.  Since the cone's refractive index $> 1$, the Gaussian pulse front slows down within the dielectric region (Fig. 11)
\begin{figure}[!h!p!b!t] \ 
\begin{center}
\includegraphics[width=5.1in]{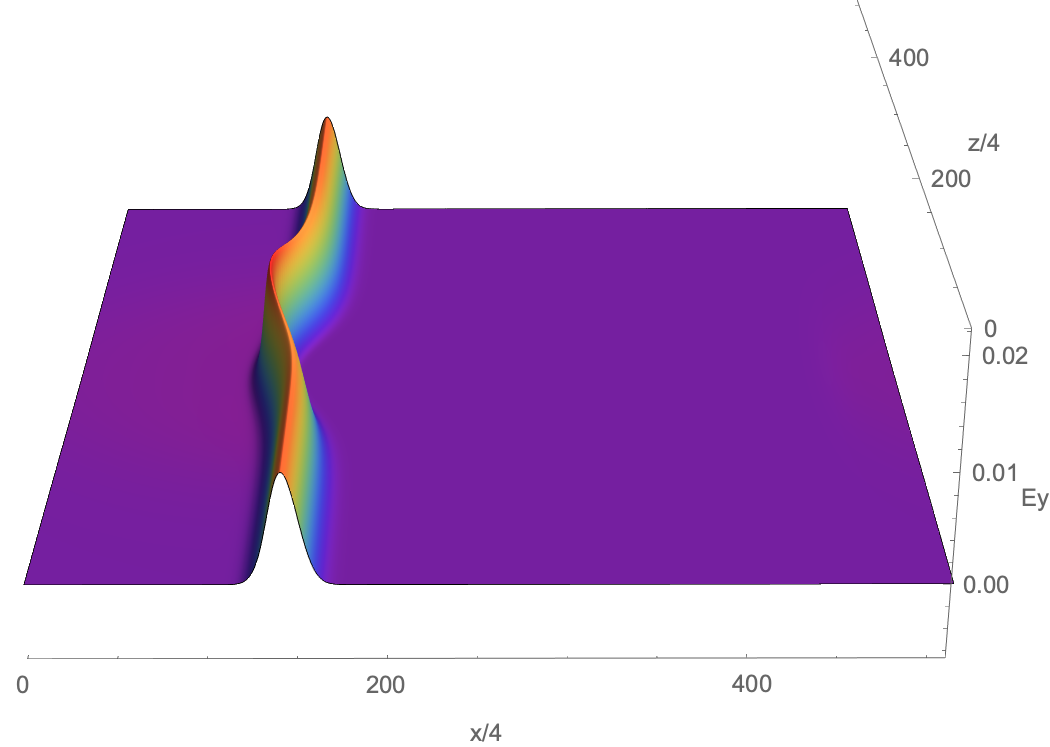}
\caption{The electromagnetic pulse as it starts to interact with the dielectric cone, $t = 1250$
}
\end{center}
\end{figure} 
\begin{figure}[!h!p!b!t] \ 
\begin{center}
\includegraphics[width=5.1in]{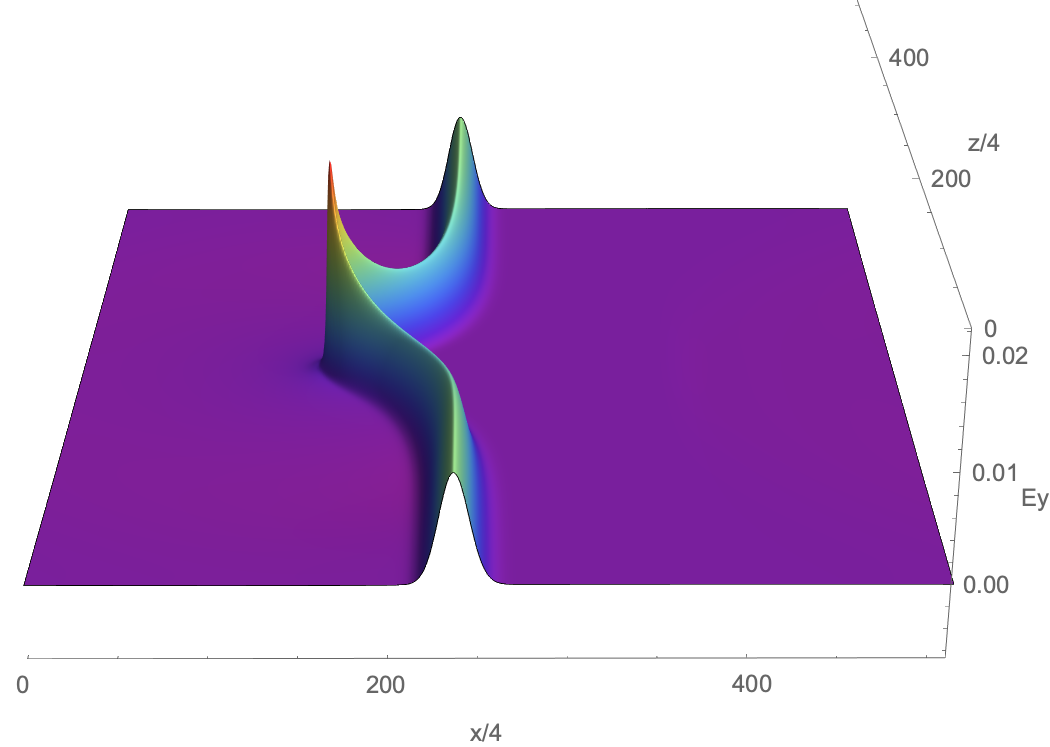}
\caption{The electromagnetic pulse at $t=2500$
}
\end{center}
\end{figure} 

\begin{figure}[!h!p!b!t] \ 
\begin{center}
\includegraphics[width=5.1in]{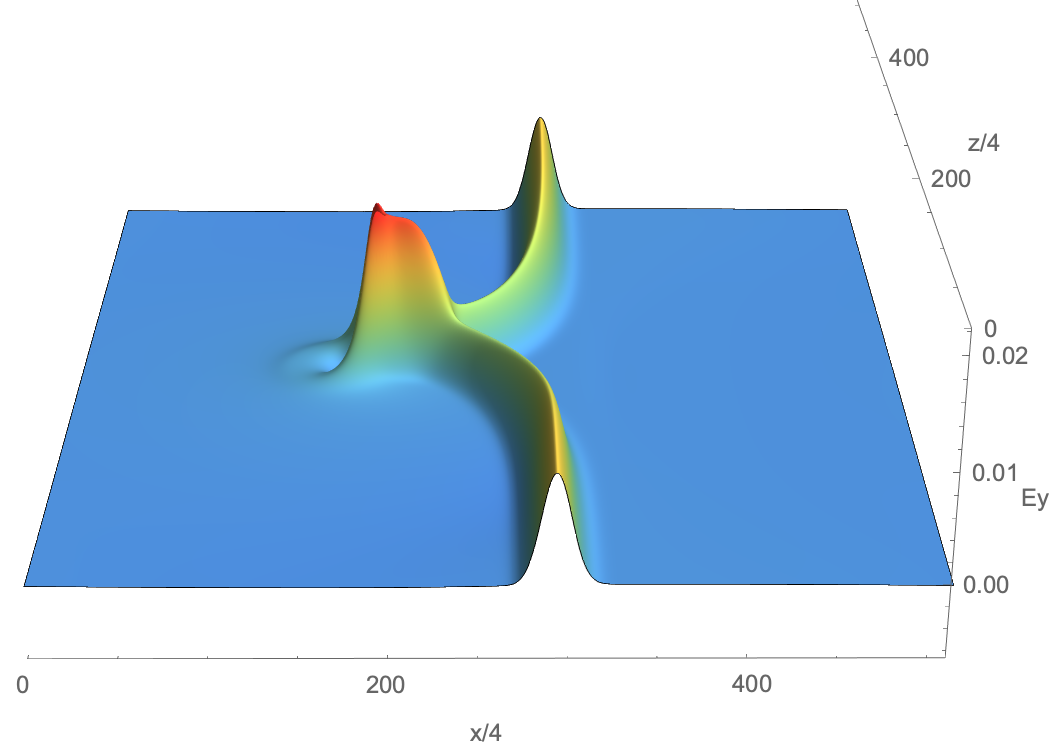}
\caption{The electromagnetic pulse at $t=3250$
}
\end{center}
\end{figure} 

\begin{figure}[!h!p!b!t] \ 
\begin{center}
\includegraphics[width=5.1in]{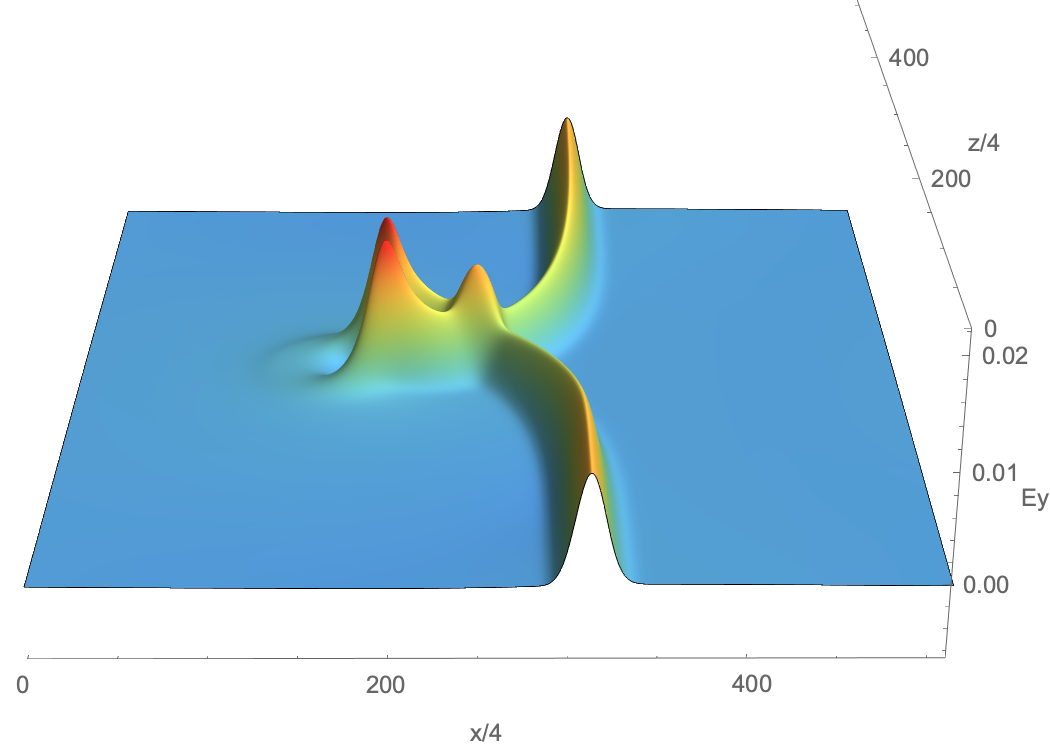}
\caption{The electromagnetic pulse at $t=3500$
}
\end{center}
\end{figure} 

\begin{figure}[!h!p!b!t] \ 
\begin{center}
\includegraphics[width=5.1in]{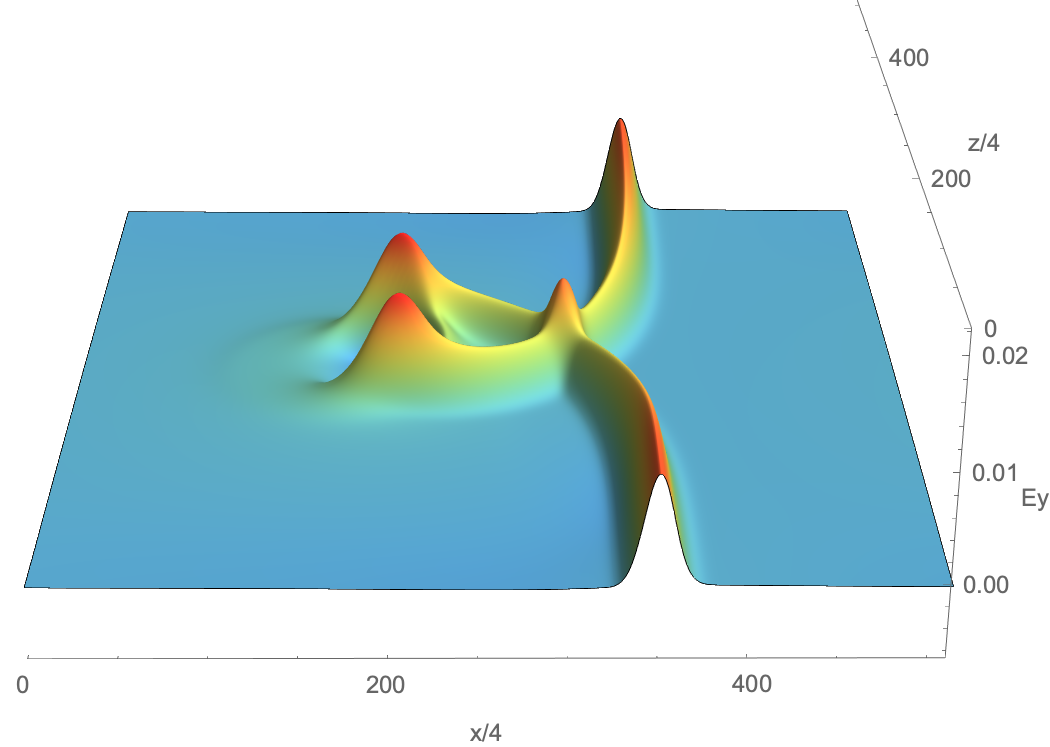}
\caption{The electromagnetic pulse at $t=4000$
}
\end{center}
\end{figure} 

\begin{figure}[!h!p!b!t] \ 
\begin{center}
\includegraphics[width=5.1in]{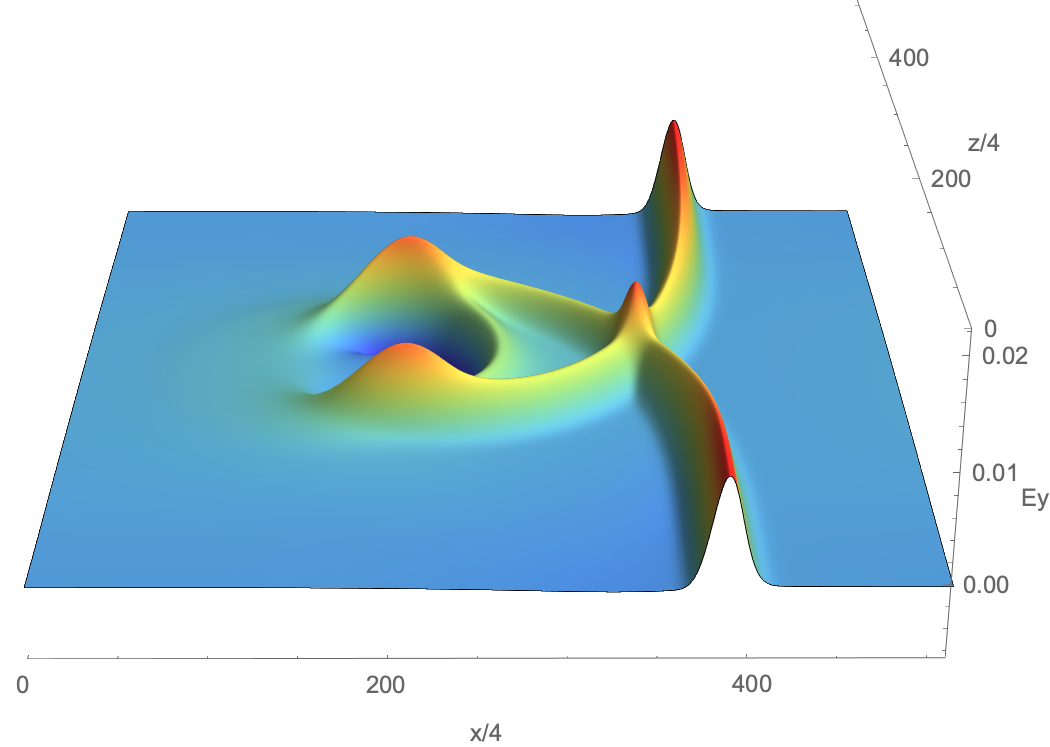}
\caption{The electromagnetic pulse at $t=4500$
}
\end{center}
\end{figure} 

\begin{figure}[!h!p!b!t] \ 
\begin{center}
\includegraphics[width=5.1in]{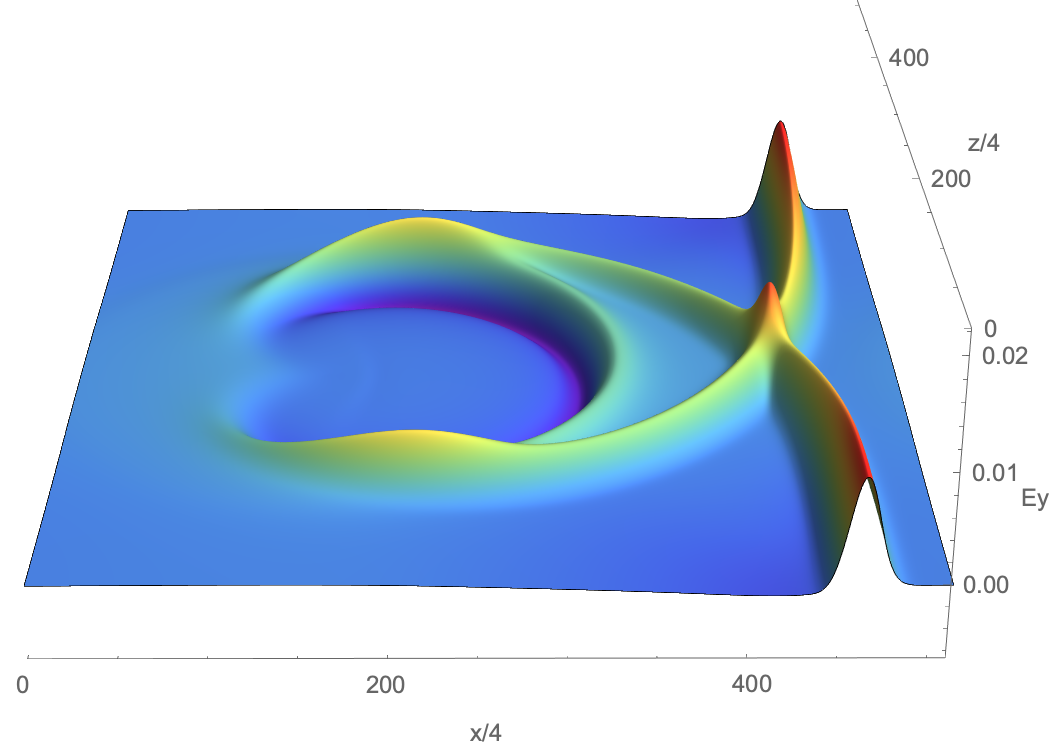}
\caption{The electromagnetic pulse at $t=5500$
}
\end{center}
\end{figure}

\section{Conclusion and Summary}
We have determined the QLA for Maxwell equations for 1D propagation in an inhomogeneous medium.  From the modular form of the Cartesian coordinates, one can readily move to 2D and to 3D inhomogenous dielectric media.  It was found that for z-propagation one required 16 qubits per lattice site because of the diagonal structure of the Pauli spin matrix $\sigma_z$.  For the non-diagonal Pauli spin matrices $\sigma_x$ and $\sigma_y$  one needs only 8 qubits per lattice site for either x-propagation or y-propagation.

\section{Acknowledgments}
 LV was partially supported by an AFRL STTR Phase I with Semicyber LLC contract number FA864919PA049.  GV, LV and MS were partially supported by an AFRL STTR Phase 2 with Semicyber LLC contract number FA864920P0419.  AKR was supported by DoE Grant Number DE-FG02-91ER-54109 and DE-SC0018090.  The 2D simulations used resources of the National Energy Research Scientific Computing Center (NERSC), a U.S. Department of Energy Office of Science User Facility operated under Contract No. DE-AC02-05CH11231, as well as the U.S. Department of Defense High Performance Supercomputer at ERDC.
 
 \section{References}
[1]  YEPEZ, J. 2002 An efficient and accurate quantum algorithm for the Dirac equation. arXiv: 0210093. 

[2]  YEPEZ, J.  2005 Relativistic Path Integral as a Lattice-Based Quantum Algorithm. Quant. Info. Proc. 4, 471-509. 

[3]  YEPEZ, J, VAHALA, G $\&$ VAHALA, L.  2009a  Vortex-antivortex pair in a Bose-Einstein condensate, Quantum lattice gas model of   theory in the mean-field approximation.  Euro. Phys. J. Special Topics 171, 9-14

[4]  YEPEZ, J, VAHALA, G, VAHALA, L $\&$ SOE, M.   2009b Superfluid turbulence from quantum Kelvin wave to classical Kolmogorov cascades.  Phys. Rev. Lett. 103, 084501. 

[5]  YEPEZ, J. 2016 Quantum lattice gas algorithmic representation of gauge field theory. SPIE 9996, paper 9996-22

[6]  OGANESOV, A, VAHALA, G, VAHALA, L, YEPEZ, J $\&$ SOE, M.  2016a. Benchmarking the Dirac-generated unitary lattice qubit collision-stream algorithm for 1D vector Manakov soliton collisions.  Computers Math. with Applic.  72, 386

[7]  OGANESOV, A, FLINT, C, VAHALA, G, VAHALA, L, YEPEZ, J $\&$ SOE, M  2016b  Imaginary time integration method using a quantum lattice gas approach.  Rad Effects Defects Solids 171, 96-102

[8]  OGANESOV, A, VAHALA, G, VAHALA, L $\&$ SOE, M.  2018. Effects of Fourier Transform on the streaming in quantum lattice gas algorithms.  Rad. Eff. Def. Solids, 173, 169-174

[9]  VAHALA, G, VAHALA, L $\&$ YEPEZ, J.  2003 Quantum lattice gas representation of some classical solitons. Phys. Lett A310, 187-196

[10]  VAHALA, G, VAHALA, L $\&$ YEPEZ, J.  2004.  Inelastic vector soliton collisions: a lattice-based quantum representation. Phil. Trans: Mathematical, Physical and Engineering Sciences, The Royal Society, 362, 1677-1690

[11]  VAHALA, G, VAHALA, L $\&$ YEPEZ, J.  2005  Quantum lattice representations for vector solitons in external potentials. Physica A362, 215-221.

[12]  VAHALA, G, YEPEZ, J, VAHALA, L, SOE, M, ZHANG, B, $\&$ ZIEGELER, S. 2011 Poincaré recurrence and spectral cascades in three-dimensional quantum turbulence.  Phys. Rev. E84, 046713

[13]  VAHALA, G, YEPEZ, J, VAHALA, L $\&$SOE, M, 2012  Unitary qubit lattice simulations of complex vortex structures.  Comput. Sci. Discovery 5, 014013

[14]  VAHALA, G, ZHANG, B, YEPEZ, J, VAHALA. L $\&$ SOE, M.  2012 Unitary Qubit Lattice Gas Representation of 2D and 3D Quantum Turbulence.  Chpt. 11 (pp. 239 - 272), in Advanced Fluid Dynamics, ed. H. W. Oh, (InTech Publishers, Croatia)

[15]  VAHALA, G, VAHALA, L $\&$ SOE, M.  2020. Qubit Unitary Lattice Algorithm for Spin-2 Bose Einstein Condensates: I – Theory and Pade Initial Conditions.  Rad. Eff. Def. Solids 175, 102-112

[16]  VAHALA, G, SOE, M $\&$ VAHALA, L.  2020  Qubit Unitary Lattice Algorithm for Spin-2 Bose Einstein Condensates: II – Vortex Reconnection Simulations and non-Abelian Vortices.  Rad. Eff. Def. Solids 175, 113-119

[17]  VAHALA, G, VAHALA, L,  SOE, M $\&$ RAM, A, K.  2020.  Unitary Quantum Lattice Simulations for Maxwell Equations in Vacuum and in Dielectric Media, arXiv: 2002.08450

[18]  VAHALA, L, VAHALA, G $\&$ YEPEZ, J. 2003  Lattice Boltzmann and quantum lattice gas representations of one-dimensional magnetohydrodynamic turbulence. Phys. Lett  A306, 227-234.

[19]  VAHALA, L, SOE, M, VAHALA, G $\&$ YEPEZ, J.  2019a. Unitary qubit lattice algorithms for spin-1 Bose-Einstein condensates.  Rad Eff. Def. Solids 174, 46-55

[20]  VAHALA, L, VAHALA, G, SOE, M, RAM, A $\&$ YEPEZ, J.  2019b. Unitary qubit lattice algorithm for three-dimensional vortex solitons in hyperbolic self-defocusing media.  Commun Nonlinear Sci  Numer Simulat 75, 152-159 

[21]  KHAN, S. A. 2005  Maxwell Optics:  I.  An exact matrix representation of the Maxwell equations in a medium.  Physica Scripta 71, 440-442;  also arXiv: 0205083v1 (2002)

[22]  CHILDS, A, N $\&$ WIEBE, N.  2012.  Hamiltonian simulation using linear combinations of unitary operations.  Quantum Info. Comput.12, 901–924.

[23]  DIRAC, P. A. M, 1928 The Quantum Theory of the Electron. Proc. Roy. Soc. A 117, 610-624.

[24]  BIALYNICKI-BIRULA, I. 1996  Photon Wave Function , in Progress in Optics, Vol. 34, pp. 248-294, ed. E. Wolf (North-Holland).

[25]  LAPORTE, O. $\&$ UHLENBECK, G. E. 1931 Application of spinor analysis to the Maxwell and Dirac    equations. Phys. Rev. 37, 1380-1397.

[26]  OPPENHEIMER, J. R. 1931 Note on light quanta and the electromagnetic field. Phys. Rev. 38, 725-746.

[27]  MOSES, E.  1959  Solutions of Maxwell’s equations in terms of a spinor notation:  the direct and inverse problems.  Phys. Rev. 113, 1670-1679

[28]  COFFEY, M, W.  2008  Quantum lattice gas approach for the Maxwell equations. Quantum Info. Processing 7, 275-281

[29]  KULYABOV, D, S, KOLOKOVA, A, V $\&$ SEVATIANOV,L , A.  2017  Spinor representation of Maxwell’s equations.  IOP Conf. Series: J. Physics: Conf. Series 788, 012025

[30]  JACKSON, J, D.  1998. “Classical Electrodynamics”, 3rd Ed., (Wiley, New York)

[31]   VAHALA, G, VAHALA, L,  SOE, M $\&$ RAM, A, K.  2020.  Unitary Quantum Lattice Simulations for Maxwell Equations in Vacuum and in Dielectric Media (to be published 2020, J. Plasma Physics

\end{document}